\documentclass[APS,preprint,superscriptaddress]{revtex4}

\usepackage{amsmath}
\usepackage{amssymb}
\usepackage{mathtools}
\usepackage{graphicx}
\renewcommand{\deg}{$^\circ$}
\newcommand{\deriv}{\mathrm{d}}
\begin{document}

\frenchspacing

\title{Thickness dependence of the resistivity of platinum-group metal thin films}

\author{Shibesh Dutta}
\affiliation{Imec, B-3001 Leuven, Belgium}
\affiliation{KU Leuven, Department of Physics and Astronomy, B-3001 Leuven, Belgium}

\author{Kiroubanand Sankaran}
\affiliation{Imec, B-3001 Leuven, Belgium}

\author{Kristof Moors}
\affiliation{Imec, B-3001 Leuven, Belgium}
\affiliation{KU Leuven, Department of Physics and Astronomy, B-3001 Leuven, Belgium}

\author{Geoffrey Pourtois}
\affiliation{Imec, B-3001 Leuven, Belgium}
\affiliation{Department of Chemistry, Plasmant Research Group, University of Antwerp, B-2610 Wilrijk-Antwerpen, Belgium}

\author{Sven Van Elshocht}
\author{J\"urgen B\"ommels}
\affiliation{Imec, B-3001 Leuven, Belgium}

\author{Wilfried Vandervorst}
\affiliation{Imec, B-3001 Leuven, Belgium}
\affiliation{KU Leuven, Department of Physics and Astronomy, B-3001 Leuven, Belgium}

\author{Zsolt T\H{o}kei}
\author{Christoph Adelmann}
\email[Author to whom correspondence should be addressed. Electronic mail: ]{christoph.adelmann@imec.be}
\affiliation{Imec, B-3001 Leuven, Belgium}

\begin{abstract}
We report on the thin film resistivity of several platinum-group metals (Ru, Pd, Ir, Pt). Platinum-group thin films show comparable or lower resistivities than Cu for film thicknesses below about 5\,nm due to a weaker thickness dependence of the resistivity. Based on experimentally determined mean linear distances between grain boundaries as well as \emph{ab initio} calculations of the electron mean free path, the data for Ru, Ir, and Cu were modeled within the semiclassical Mayadas--Shatzkes model [Phys. Rev. B \textbf{1}, 1382 (1970)] to assess the combined contributions of surface and grain boundary scattering to the resistivity. For Ru, the modeling results indicated that surface scattering was strongly dependent on the surrounding material with nearly specular scattering at interfaces with SiO$_2$ or air but with diffuse scattering at interfaces with TaN. The dependence of the thin film resistivity on the mean free path is also discussed within the Mayadas--Shatzkes model in consideration of the experimental findings. 
\end{abstract} 

\maketitle

\section{Introduction}
\noindent Finite size effects in the resistivity of metallic thin films or nanowires have been a topic of research for several decades both from a fundamental as well as an applied point of view. While the resistivity of bulk metals is dominated by phonon (and possibly impurity) scattering at room temperature, surface scattering can become dominant when the size of the thin films or nanowires is reduced \cite{1,2,ZB}. In addition, grain sizes (\textit{i.e.} mean linear distances between grain boundaries) in polycrystalline films or wires have typically been found to decrease for decreasing film thickness or wire diameter, which leads to an increasing contribution of grain boundary scattering in thin films or nanowires \cite{3,4}. Ultimately, when the structure size becomes of the order of a few nanometer, electron confinement effects will also alter the resistivity of metallic nanostructures \cite{Tesa,NA,ZB,Mey,Moors,Hedge1,Jones,Lanzillo,Munoz}. While this behavior is universally found in all metals, there is still controversy over the relative importance of the different additional scattering contributions even for the most studied material, Cu \cite{7,8,Maitre,Marom,Josell,Sun1,Graham,Sun2,Chawla1,Chawla3}, and only few comparative studies for different metals have been reported \cite{DeVries,Tay,Camacho}. 

From an applied point of view, the understanding of the resistivity of metals in small dimensions is crucial since metallic nanowires form the interconnect structures that are used in integrated microelectronic circuits. At present, the widths of scaled interconnect wires are of the order of 25 to 30\,nm and are expected to reach dimensions of about 10\,nm in the next decade. At such dimensions, surface and grain boundary scattering in Cu, the standard conductor material presently used in interconnects, dominate over phonon scattering, resulting in resistivities much larger than in the bulk \cite{5,6,7,8,Maitre} and leading to a deterioration of the interconnect properties \cite{9,10,11,Ceyhan}. In addition, Cu-based interconnects require diffusion barriers and adhesion liners to ensure their reliability. Since their resistivity is typically much higher than that of Cu, their contribution to the wire conductance is often negligible. Barriers and liners are difficult to scale and may occupy a significant volume when the interconnect width approaches 10\,nm, reducing the volume available for Cu. Therefore, alternative metals have recently elicited much interest as they could serve as a barrierless replacement for Cu. Among them, platinum-group metals (PGMs) have emerged as promising candidates due to the combination of low bulk resistivity, resistance to oxidation, and high melting point, which can be considered as a proxy for resistance to electromigration \cite{12,13,Kirou2,14}.

The main quest for alternatives to Cu is motivated by the observation that the resistivity increase for thin films due to surface or grain boundary scattering depends on $\lambda/\ell$. Here, $\lambda$ is the intrinsic mean free path (MFP) of the charge carriers in the metal and the characteristic length scale $\ell$ is the film thickness for surface scattering or the average linear distance between grain boundaries for grain boundary scattering \cite{1,2,ZB,3,4}. Hence, metals with short MFPs should be inherently less sensitive to surface or grain boundary scattering for a given $\ell$. As a result, such metals may show lower resistivities than Cu for sufficiently small dimensions despite their larger bulk resistivity \cite{5,Zhang,Naeemi}. In addition, quantum effects for very small $\ell$ may also lead to such a behavior \cite{Simbeck,Gao,Lanzillo}. However, such a crossover behavior has been elusive so far despite its strong interest for interconnect metallization.

In this paper, we discuss the thickness dependence of the resistivity of PGM (Ru, Pd, Ir, Pt) ultrathin films with thicknesses between 3 and 30\,nm. We demonstrate that their resistivity exhibits a much weaker thickness dependence than that of Cu films in the same thickness range. As a result, for films thinner than about 5\,nm, the resistivities of Ru and Ir films fall below that of Cu. The thickness dependence of the resistivity of Ru and Ir is then modeled using the analytical semiclassical Mayadas--Shatzkes approach. We demonstrate that, within the Mayadas--Shatzkes model, the shorter MFP of PGMs is indeed predominantly responsible for the resistivity crossover with respect to Cu. The data suggest that Ru, and Ir are promising metals for future interconnects in advanced technology nodes with interconnect widths below 10\,nm \cite{ACSRuIC}.

\section{Experimental and theoretical methods}

\noindent All films were deposited by physical vapor deposition (PVD) at room temperature on Si (100). Prior to metal deposition, a 90\,nm thick thermal oxide was grown on the Si wafers to ensure electrical isolation. Cu, Ru, and Ir films were deposited on 300 mm wafers in a Canon Anelva EC7800 system. In addition to films directly deposited on SiO$_2$, Cu and Ru films were also grown \textit{in situ} on 1.5\,nm thick PVD TaN and capped by 1.5\,nm thick PVD TaN. This was done to prevent the oxidation of the Cu surface, to avoid Cu diffusion into the underlying SiO$_2$ during annealing, and to study the effect of the ``cladding'' material on the Ru thin film resistivity. Pt films were sputter deposited on small samples in a home build system using thin ($\approx 2$\,nm) Ti to promote adhesion. Pd films were obtained by e-beam evaporation using a Pfeiffer PLS 580 system on a thin TiN adhesion layer. 

The thin film resistivity was calculated from sheet resistance measurements using KLA Tencor RS100 and Jandel 4-point probe systems as well as the film thickness measured by x-ray reflectivity (XRR). XRR was performed using Cu K$\alpha$ radiation in a Bede MetrixL diffractometer from Jordan Valley or a Panalytical X'Pert diffractometer. Film thicknesses by XRR were cross-calibrated by Rutherford backscattering spectrometry (RBS) measurements using a 1.52 MeV He$^+$ ion beam in a rotating random mode at a backscatter angle of 170\deg . In all cases, the contribution of the adhesion layers to the sheet resistance was obtained by independent measurements and taken into account in the determination of the PGM thin film resistivity. The diffractometers mentioned above were also used to assess the film crystallinity using x-ray diffraction (XRD). Surface roughnesses were measured by atomic force microscopy (AFM) using a Bruker IconPT microscope. Lateral correlation lengths of the surface roughness were obtained from the autocorrelation function. The microstructure of the films was determined from plan-view transmission electron microscopy (TEM) images using Tecnai F30 and Titan3 G2 microscopes. Based on these images, the mean linear grain boundary intercept distance (the average linear distance between grain boundaries) was determined \cite{ASTM,Abrams}. Due to the almost columnar nature of the microstructure and their expected relatively weak contribution to the resistivity, grain boundaries parallel to the surface were neglected. 

Electronic structures of the PGMs, Ru, Rh, Pd, Os, Ir, and Pt, as well as of Cu were obtained by first-principles calculations based on density functional theory as implemented in the Quantum {\sc Espresso} package \cite{K8}. Projector augmented wave \cite{K9} potentials with the Perdew--Burke--Ernzerhof generalized gradient \cite{K10} approximation of the exchange-correlation functional have been used together with a $40\times 40\times 40$ Monkhorst--Pack \textbf{k}-point sampling grid and an energy cutoff of 80\,Ry to ensure the convergence of the total energy differences (10 -- 12\,eV). The Fermi surface $S_{F,n}(\textbf{k})$ was determined from the calculated electron energy as a function of the wave vector \textbf{k} for each band with index $n$. In addition, using the obtained Fermi surfaces and electronic densities of states, scattering times due to electron--phonon interactions have been calculated using standard first-order perturbation theory \cite{Allen}. Details of these calculations can be found in Ref.~\cite{13}.

\section{Material properties and resistivity of platinum-group metal thin films}

\begin{figure}[p]
\includegraphics[width=8.5cm]{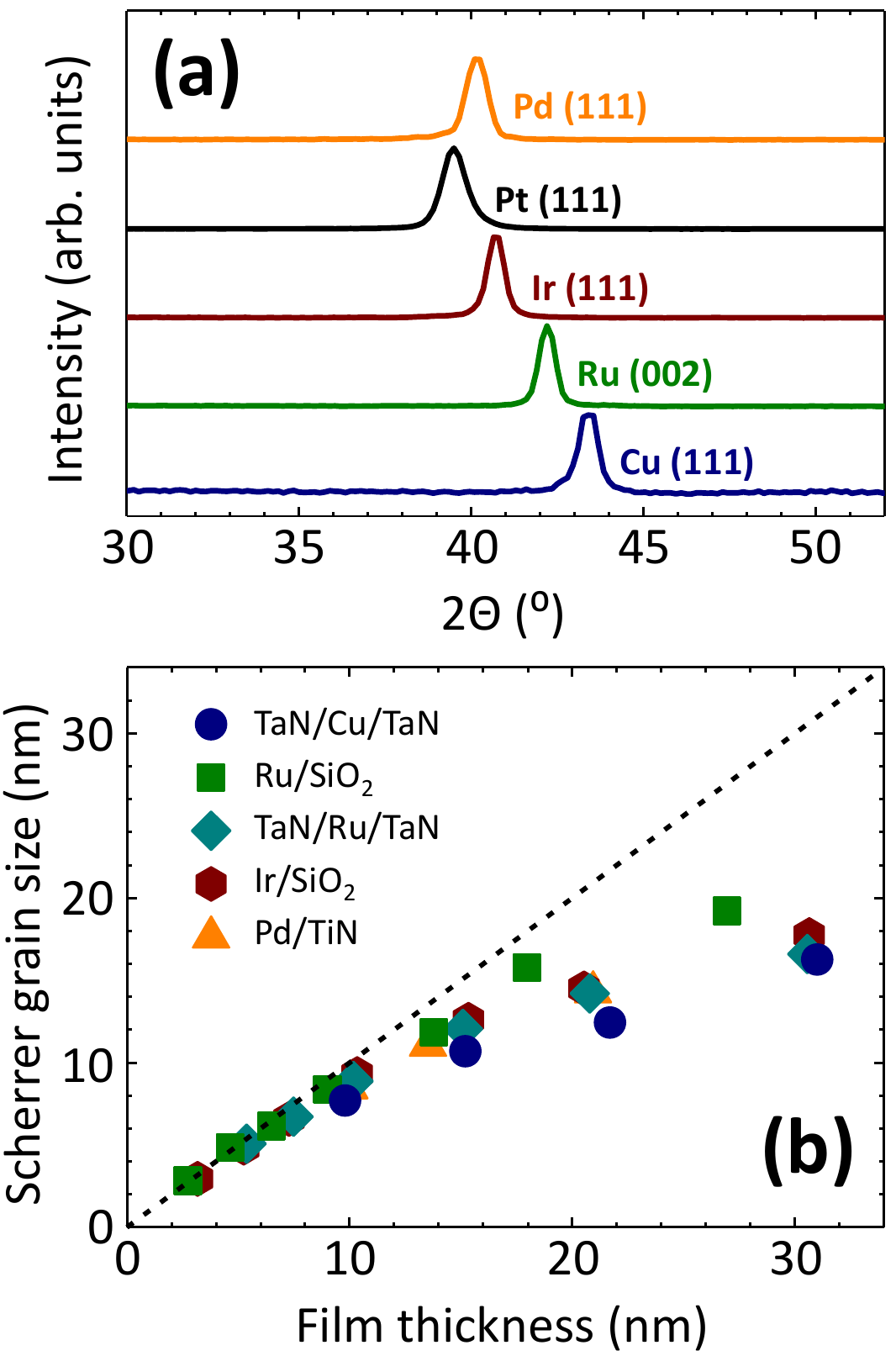} 
\caption{\label{XRD} (a) $\Theta$-$2\Theta$ XRD pattern of 20\,nm thick films of platinum-group metals and Cu, as indicated (Ru was deposited on SiO$_2$). All layers have been annealed at 420\,\deg C for 20\,min. The patterns indicate strong (111) texture for fcc materials (Pd, Pt, Ir, Cu) and (001) texture for hcp Ru. (b) Out-of-plane Scherrer grain (crystallite) size determined from the XRD pattern. The dashed line indicates the expected behavior for ideal columnar growth, \textit{i.e.} for crystallite sizes equal to the film thickness.}
\end{figure}

\noindent All films were polycrystalline as deposited. The $\Theta$-$2\Theta$ XRD patterns were consistent with the expected crystal structures of the stable phases (hcp for Ru, fcc for all other metals) with (partial) texture [(001) for Ru, (111) for other metals]. Post deposition annealing at 420\,\deg C in forming gas for 20\,min improved both the crystallinity and led to strong texturing [Fig.~\ref{XRD}(a)]. The out-of-plane Scherrer crystallite size of annealed films [Fig~\ref{XRD}(b)] was typically of the order of the film thickness for films up to about 15\,nm and deviated slightly towards smaller values for thicker films. This indicates that the microstructure of the films was (nearly) columnar.

\begin{figure}[p]
\includegraphics[width=8.5cm]{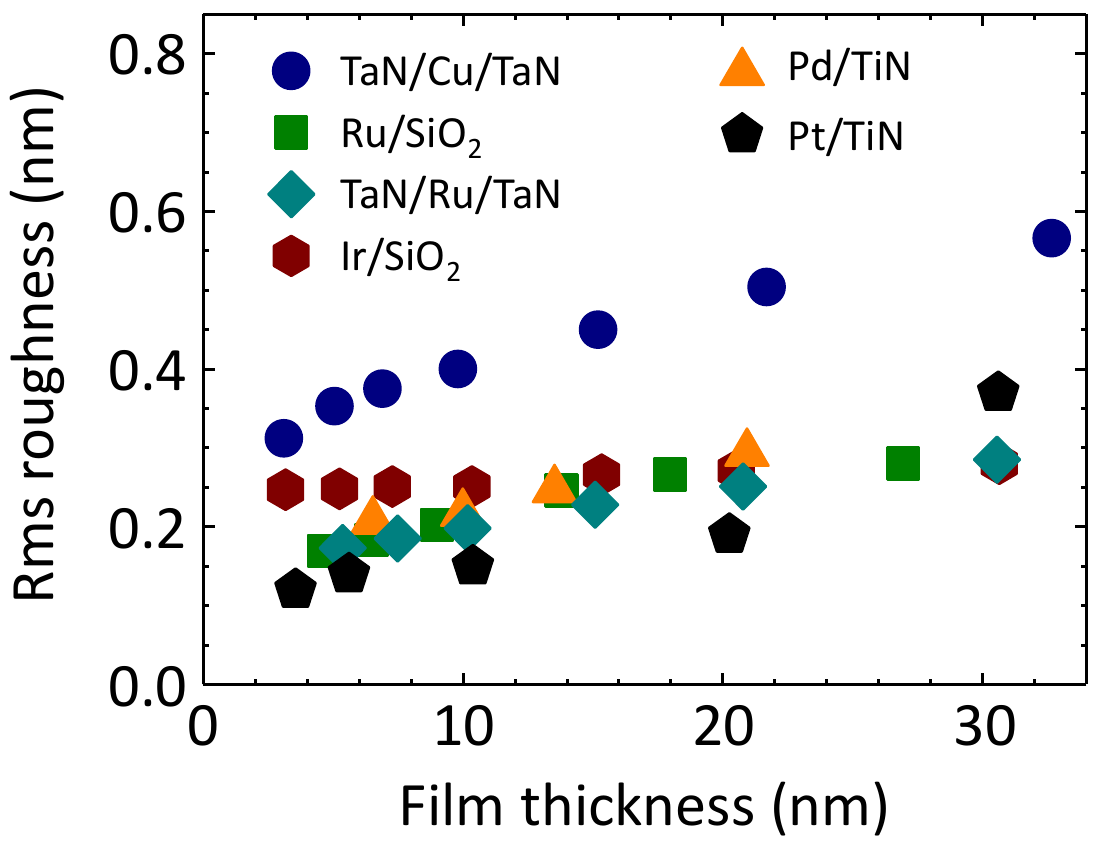} 
\caption{\label{AFM} RMS surface roughness of platinum-group metal and Cu thin films as a function of their thickness.}
\end{figure}

Figure~\ref{AFM} shows the root-mean-square (RMS) roughness of Cu (TaN/Cu/TaN), Ru (both Ru/SiO$_2$ and TaN/Ru/TaN), Ir, Pd, and Pt films as a function of the film thickness after post deposition annealing at 420\deg C. The roughness of the annealed films increased with increasing film thickness but remained below 0.4\,nm even for 30\,nm thick PGM films. Cu films were slightly rougher with RMS values of 0.5 to 0.6\,nm for the thickest films. XRR measurements (not shown) indicated that the roughness of the top surface was very similar to that of buried interfaces (typically also 0.3 to 0.5\,nm). The lateral correlation length $\xi$ of the surface roughness (obtained with Gaussian correlation statistics) was between 10 and 15\,nm for all films with insignificant differences between materials/stacks and only little coarsening in the studied thickness range up to 30\,nm. 

\begin{figure}[p]
\includegraphics[width=8cm]{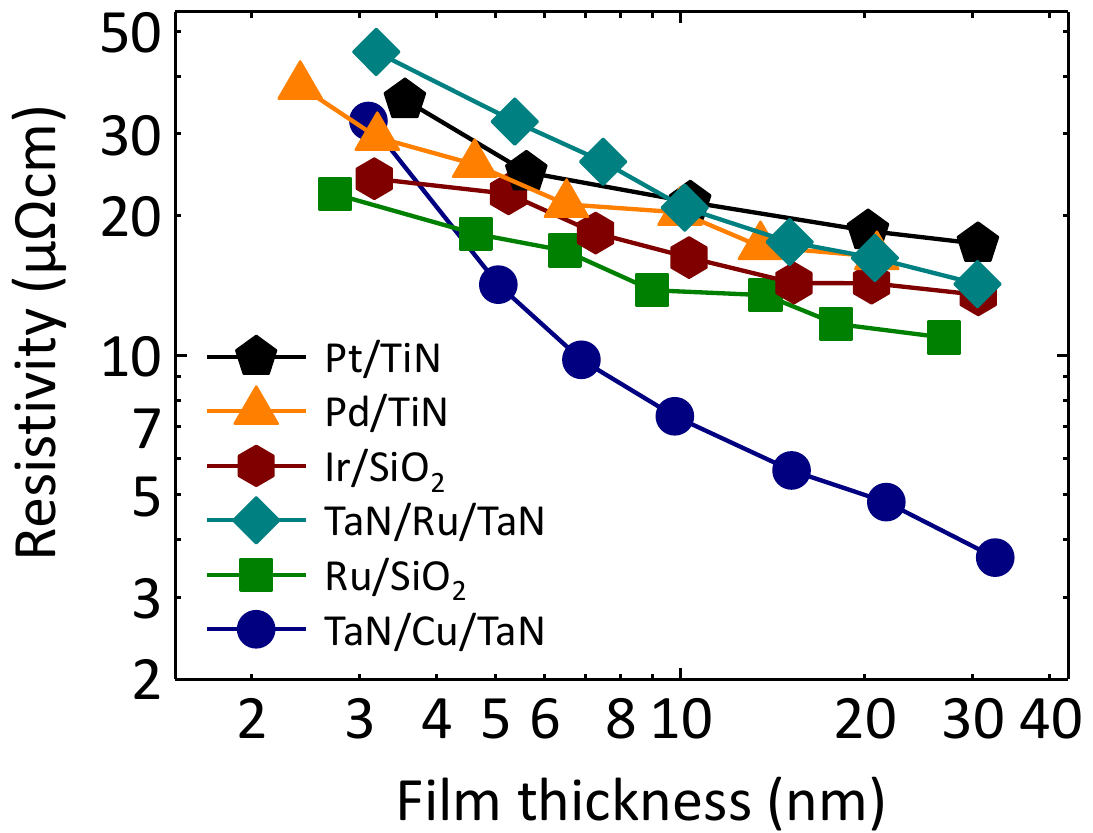} 
\caption{\label{TF_Res} Thickness dependence of the thin film resistivity of platinum-group metals and Cu, as indicated. All films have been annealed at 420\,\deg C for 20\,min.}
\end{figure}

Figure~\ref{TF_Res} shows the resistivity of Cu (TaN/Cu/TaN), Ru (Ru/SiO$_2$ and TaN/Ru/TaN), Pd, Ir, and Pt as a function of the film thickness. All films were annealed at 420\,\deg C for 20\,min in forming gas. Cu showed a strong increase with decreasing film thickness, as observed previously and ascribed to the combination of surface and grain boundary scattering. Note that the Cu resistivity values were close to the ones reported in the literature for scaled Cu interconnect lines of the same critical dimension \cite{ITRS,PR}.

By contrast, all PGM thin films showed a much weaker thickness dependence of the resistivity than Cu. For films with thicknesses of 10\,nm and above, the resistivities were much higher than for Cu owing to the higher bulk resistivities of PGMs. However, for film thicknesses of about 5\,nm and below, the resistivities of Ru and Ir became comparable and even lower than the resistivity of Cu. From a technological point of view, this resistivity crossover renders PGMs, in particular Ru and Ir highly interesting for scaled interconnects with critical dimensions of 10\,nm and below, where the current combination of Cu, diffusion barriers, and adhesion liner layers may be outperformed by barrierless Ru or Ir metallization. Indeed, scaled Ru filled interconnect structures have already shown first promising results \cite{ACSRuIC} demonstrating the prospects of these materials for future interconnect technology nodes. 

From a more fundamental point of view, these data raise the question of the material dependence of the thin film scattering contributions, such as surface and grain boundary scattering. It has been asserted that a shorter electron MFP leads to a weaker thickness dependence of both surface and grain boundary scattering \cite{5,Zhang,Naeemi}. However, to confirm this argument, effects of potentially different microstructures (\emph{e.g.} the thickness dependence of the mean linear distance between grain boundaries) have to be understood. 

\section{Fermi surfaces and electron mean free paths of platinum-group metals} 

In a first step, we have computed the bulk electron MFPs of the PGMs as well as of Cu. The MFP in transport direction \textbf{t} of an electron with wave vector \textbf{k} is given by $\lambda_{n,\textbf{t}}(\textbf{k}) = \tau_n\left(\textbf{k}\right)\times \left|\textbf{v}_{n,\textbf{t}}(\textbf{k})\right|$, where $\tau_n(\textbf{k})$ is the relaxation time of an electron with wave vector \textbf{k} and $\textbf{v}_{n, \textbf{t}}(\textbf{k})$ is the projection of the Fermi velocity $\textbf{v}_{n}(\textbf{k})= (1/\hbar)\nabla_\textbf{k} E_{F,n}$ on the transport direction. Here, $E_{F,n}$ is the Fermi energy of the band with index $n$. 

In a semiclassical approximation, the conductivity along the transport direction \textbf{t} can be expressed as \cite{Jacoboni,Mahan}

\begin{equation}
\sigma_\textbf{t} = -2 \frac{e^2}{(2\pi)^3} \sum_n \int \deriv^3 \textbf{k} \; \left|\textbf{v}_{n, \textbf{t}}(\textbf{k})\right|^2 \tau_n(\textbf{k})
		\frac{\partial f_n(\textbf{k})}{\partial \epsilon},
\end{equation}

\noindent where $f_n(\textbf{k})$ is the (Fermi) distribution function, and $\epsilon$ has the dimension of an energy. The summation is carried out over the band index $n$. At low temperature, ${\partial f_n(\textbf{k}})/{\partial \epsilon} = - \delta (\epsilon_n(\textbf{k})- E_F)$. Assuming that the relaxation time is isotropic and does not depend on the band index, \emph{i.e.} $\tau_n(\textbf{k}) \equiv \tau$, one obtains \cite{14}

\begin{equation}
\label{lambda}
\frac{\sigma_\textbf{t}}{\tau} = \frac{1}{\tau\rho_{\textbf{t}}} = \frac{e^2}{4\pi^3\hbar}\sum_n\int_{S_{F,n}}\deriv S \; \frac{\left|\textbf{v}_{n,\textbf{t}}(\textbf{k})\right|^2}{\left|\textbf{v}_{n}(\textbf{k})\right|}.
\end{equation}

\noindent Here, the integration is carried out over the Fermi surface. Hence, the product of the relaxation time and the bulk resistivity depends only on the morphology of the Fermi surface. When the bulk resistivity is known, $\tau$ can be deduced and thus the MFP in the transport direction, $\lambda_{n,\textbf{t}}(\textbf{k}) = \tau\times \left|\textbf{v}_{n,\textbf{t}}(\textbf{k})\right|$. For polycrystalline materials, suitable averages can be obtained from the isotropic $\tau$ in combination with an average of $\textbf{v}_{n,\textbf{t}}(\textbf{k})$ over the relevant transport directions.

\begin{figure*}[p]
\includegraphics[width=17cm]{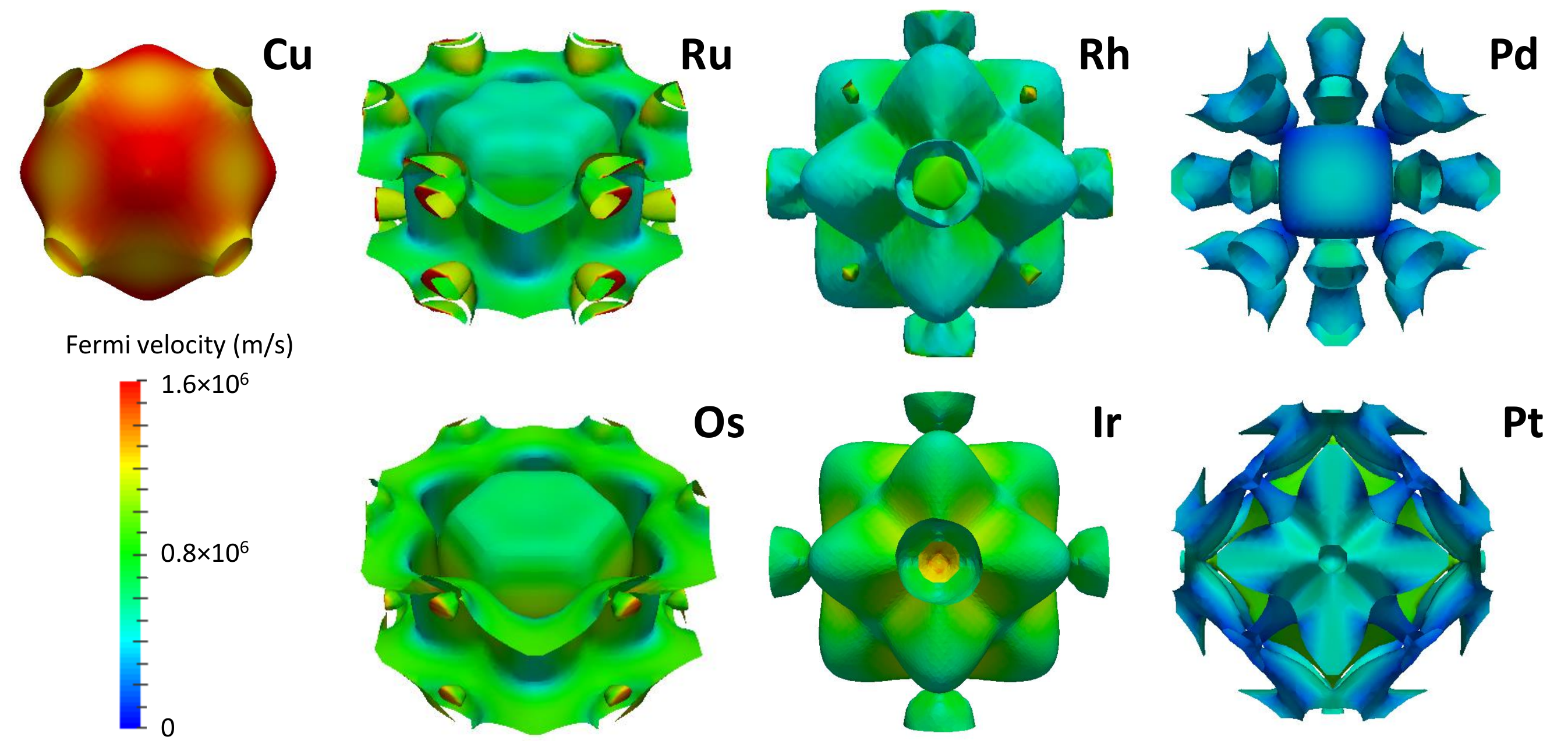} 
\caption{\label{Fermi} Fermi surfaces of platinum-group metals. The Fermi surface of Cu is also shown as a reference. The color scheme indicates the Fermi velocity.}
\end{figure*}

Using \textit{ab initio} calculated Fermi surfaces (Fig.~\ref{Fermi}), $\tau\times\rho$ was calculated for all PGMs as well as for Cu as a reference. Based on these values and experimental bulk resistivities $\rho_0$ \cite{Resistivity}, relaxation times $\tau_0$ were then deduced. In addition, relaxation times $\tau_c$ due to electron--phonon scattering were directly calculated \cite{13,Allen}. Since all films were polycrystalline and textured, an effective Fermi velocity $v_\mathrm{ave}$ was obtained by averaging $\textbf{v}_{n,\textbf{t}}(\textbf{k})$ over transport directions perpendicular to [001] for the hcp metals (Ru, Os) and [111] for the fcc metals (Rh, Pd, Ir, Pt, and Cu). MFP values $\lambda_0$ and $\lambda_c$ were then calculated using $\tau_0$ and $\tau_c$, respectively. The results are summarized in Tab.~\ref{Tab_Ab_Initio}. Note that the resistivity of bulk hcp Ru and Os is anisotropic with the higher resistivity perpendicular to [001] \cite{Ru_aniso}, \emph{i.e.} along the transport direction of our textured films. The values are generally in good agreement with a previous report \cite{14}. 

As discussed above, metals with short MFPs may be less sensitive to surface or grain boundary scattering and thus may show a weaker thickness dependence of the resistivity than Cu \cite{5,Zhang,Naeemi}. As indicated in Tab.~\ref{Tab_Ab_Initio}, all PGMs show significantly shorter MFPs than Cu. Since the thin film resistivity for a given thickness or linear grain boundary distance also depends on the bulk resistivity, $(\lambda\rho_0)^{-1}$ has been used as a figure of merit of a metal for the expected resistivity scaling at small dimensions \cite{13,Kirou2,14}. As shown in Tab.~\ref{Tab_Ab_Initio}, all PGMs show higher figures of merit than Cu with Pt showing the highest value. However, due to a comparatively high bulk resistivity, Pt (as well as Pd and Os) may show benefits only for very small thicknesses or short linear grain boundary distances.

\section{Semiclassical thin film resistivity modeling} 

\noindent To gain further insight into the contributions of surface and grain boundary scattering, the resistivity of Ru, Ir and Cu was modeled using the semiclassical model developed by Mayadas and Shatzkes \cite{4}. Despite recent advances in ab initio modeling \cite{Hedge1, Hedge2, Zhou, Jones, Lanzillo}, the approach by Mayadas and Shatzkes remains the only tractable quantitative model for thin film resistivities in the studied thickness range up to 30\,nm that contains both surface and grain boundary scattering. Here, transport is calculated within a Boltzmann framework using an isotropic Fermi surface. Band structure effects are however included in our calculations via an anisotropic mean free path, as discussed above. 

The model also neglects confinement effects that are expected to further increase the resistivity. For nanowires, \textit{ab initio} calculations have shown an orientation dependent increase of the resistivity although the magnitude of the increase varied between studies \cite{Hedge1,Hedge2,Jones,Lanzillo,Zhou}. As a consequence, the thickness dependence of confinement effects in nanowires (and the transition to bulk-like behavior) cannot be considered as fully understood for (Cu) nanowires and even less so for (Cu or PGM) thin films, where confinement effects are expected to be weaker than for nanowires. Estimations of confinement effects within an anisotropic effective mass approximation (using anisotropic effective masses calculated from the above Fermi surfaces) lead to characteristic energies of the order of a few ($< 15$) meV for 5\,nm thick Cu, Ru, and Ir films and the experimentally observed textures. These confinement energies are much smaller than the Fermi energy (and hence a large number of subbands are occupied) and even $k_BT$ at room temperature. Recent \textit{ab initio} results by Zhou \textit{et al.} \cite{Zhou} and Lanzillo \cite{Lanzillo} for Cu suggest that expected confinement effects are still small compared to the experimentally observed increase of the thin film resistivity with respect to the bulk; therefore, grain boundary and surface scattering are expected to dominate over confinement effects in our thin films. We also note that we have observed no different trends for the thinnest films of 5\,nm thickness and below, in the sense that fitting data subsets including thicker films only did not lead to significantly different fitting parameters. However, future work will be required to unambiguously identify the effect of band structure and confinement on the thin film resistivity, especially for film thicknesses far below 10\,nm.

In the Mayadas--Shatzkes model, the resistivity of a thin film with thickness $h$ and average linear distance between grain boundaries (average linear intercept length \cite{RemGS}) $l$ is given by

\begin{equation}
\begin{split}
\label{MSS}
\rho_{tf} = \left[ \frac{1}{\rho_\mathrm{GB}} - \frac{6}{\pi \kappa \rho_0}\left( 1-p\right)\int\limits_0^{\pi/2} d\phi \int\limits_1^\infty dt \frac{\cos^2\phi}{H^2} \times\right. \\ \left. \left(\frac{1}{t^3}-\frac{1}{t^5}\right) \frac{1-e^{-\kappa tH}}{1-pe^{-\kappa tH}} \right] ^{-1} \equiv \left[\frac{1}{\rho_\mathrm{GB}} - \frac{1}{\rho_\mathrm{SS,GB}}\right] ^{-1},
\end{split}
\end{equation}

\noindent with $\rho_\mathrm{GB} = \rho_0 \left[1-3\alpha /2 + 3\alpha^2 -3\alpha^3\ln\left( 1 + 1/\alpha\right) \right]^{-1}$, $H = 1 + \alpha /\cos\phi\sqrt{\left( 1-1/t^2\right)}$, $\kappa = h/\lambda$, and $\alpha = \left(\lambda /l\right) \times 2R\left(1-R\right)^{-1}$ \cite{RemR}. $p$ and $R$ are parameters that describe the surface and grain boundary scattering processes, respectively. The phenomenological surface specularity parameter $p$ varies between 0 for diffuse and 1 for specular scattering of charge carriers at the surface or interface; $R$ is the reflection coefficient ($0 < R < 1$) of a charge carrier at a grain boundary. In general, $p$ can take different values at the top and bottom interface, \emph{e.g.} when the surface and interface roughnesses are strongly different, as described by the model of Soffer \cite{Soffer}. However, given the observation that surface and buried interface roughnesses in the stacks considered here are low and very similar, we will assume that a single parameter $p$ can describe both interfaces of the metal films. 

\subsection{Film thickness dependence of the linear grain boundary distance} 

While surface scattering depends directly on the film thickness, grain boundary scattering depends on the average linear distance between grain boundaries, $l$, along the transport direction. Therefore, a quantitative model of the thin film resistivity as a function of film thickness requires the knowledge of the thickness dependence of $l$ in polycrystalline films. Historically, it has often been assumed that $l$ is identical or proportional to the film thickness and this assumption has often been used to model the thickness dependence of the resistivity. However, it has been pointed out that such simple relations are generally not valid \cite{Marom_GS}. 

\begin{figure*}[p]
\includegraphics[width=17cm]{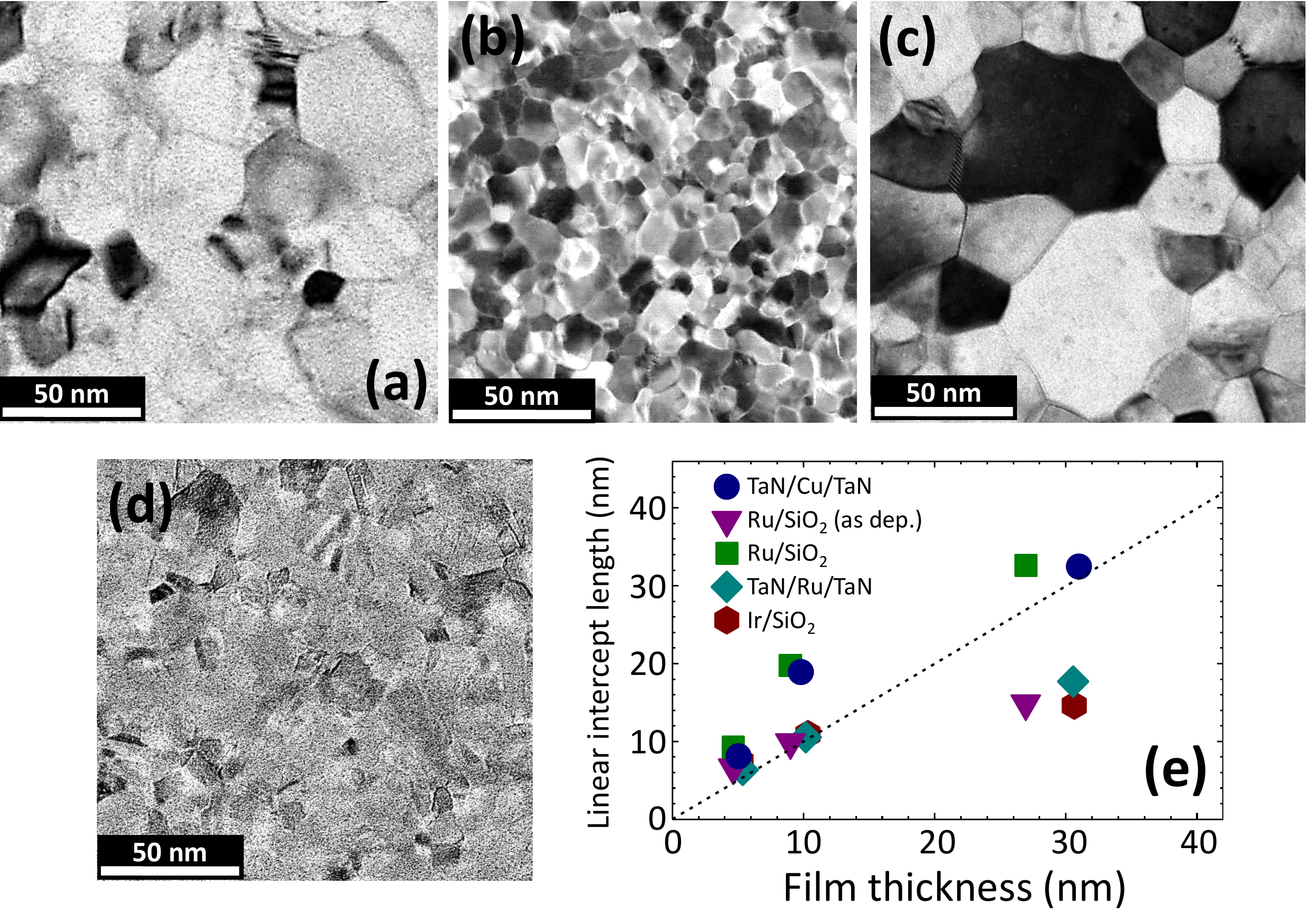} 
\caption{\label{Grain_Size} Plan-view TEM images of 30\,nm thick films of (a) TaN/Cu/TaN, (b) TaN/Ru/TaN, (c) Ru/SiO$_2$, and (d) Ir/SiO$_2$. All films have been annealed at 420\,\deg C for 20 min. (e) Grain size distributions of Ru/SiO$_2$ for both annealed and as deposited films with thicknesses of 5, 10, and 30\,nm, as indicated. (e) Mean linear intercept length between grain boundaries deduced from the TEM images \emph{vs.} film thickness. The dashed line represents the case where the linear intercept length is equal to the film thickness.}
\end{figure*}

We have therefore experimentally determined the average linear grain boundary distance using the intercept method \cite{ASTM,Abrams} from plan-view transmission electron micrographs for 5, 10, and 30\,nm thick Ru/SiO$_2$ (as deposited and after annealing at 420\,\deg C), TaN/Ru/TaN (annealed), Ir/SiO$_2$ (annealed), and TaN/Cu/TaN (annealed) thin films. Figure~\ref{Grain_Size} shows both sample TEM images as well as the deduced film thickness dependence of the mean linear grain boundary intercept length. While for TaN/Cu/TaN and annealed Ru/SiO$_2$ the mean linear intercept length was close to the film thickness, other stacks clearly showed a saturating effect for $\sim 30$\,nm thick films. Linear intercept lengths for in-between thicknesses were obtained by piecewise linear interpolation. Other more nonlinear interpolation schemes did not have any significant effects on the modeling results discussed below.

\subsection{Semiclassical modeling results and discussion}

\begin{figure}[p]
\includegraphics[width=8.5cm]{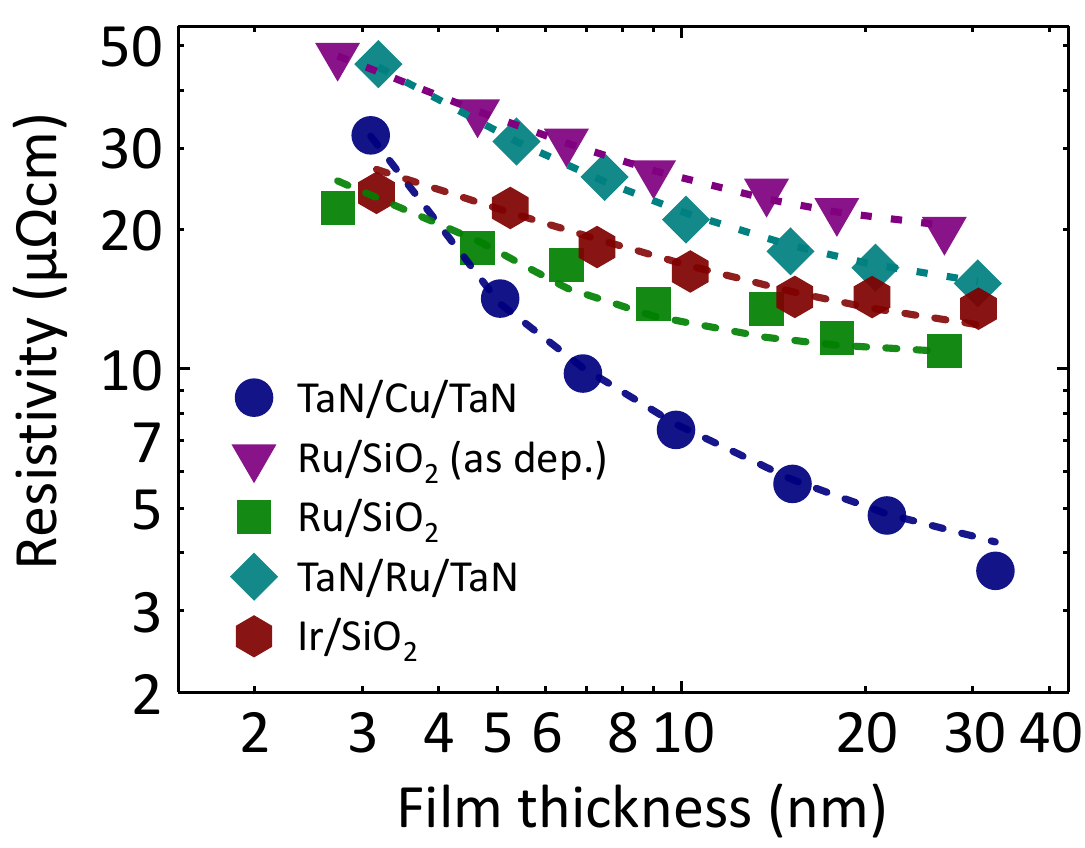} 
\caption{\label{Fits} Best fits (dashed lines) using the Mayadas--Shatzkes model [see Eq.~(\ref{MSS})] to the experimental thickness dependence (symbols) of the resistivity of platinum-group metal and Cu films, as indicated. The resulting fit parameters are listed in Tab.~\ref{Tab1}. All stacks have been annealed at 420\,\deg C except as deposited Ru/SiO$_2$.}
\end{figure}

Figure \ref{Fits} shows the experimental thickness dependence of the resistivities of Ru/SiO$_2$, TaN/Ru/TaN, Ir/SiO$_2$, and TaN/Cu/TaN, together with the best fits using the Mayadas--Shatzkes model in Eq.~(\ref{MSS}). All films were annealed at 420\,\deg C for 20\,min except for an additional data set of as deposited Ru/SiO$_2$. To obtain best fits, the experimentally determined thickness dependences of the linear distance between grain boundaries for the different materials and stacks, as discussed in the previous section, were used. In addition, bulk electron MFPs obtained by \textit{ab initio} calculations, $\lambda_0$ (see Tab.~\ref{Tab_Ab_Initio}), were employed in combination with experimental bulk resistivities. As discussed above, quantum confinement effects are difficult to quantify and have been neglected as they can still be expected to be small for the studied film thicknesses at room temperature except maybe for the very thinnest films (see also the discussion in Ref.~\cite{Choi}). For Ru, the bulk resistivity has been reported to be anisotropic \cite{Ru_aniso}. Since all films studied here showed strong (001) texture, the bulk in-plane resistivity (perpendicular to the hexagonal axis) has been used. Only $p$ and $R$ were used as adjustable parameters. In general, the model described well the thickness dependence of the thin film resistivity for all materials and stacks over the entire thickness range. The resulting parameters are listed in Tab.~\ref{Tab1}. 

Using the Mayadas--Shatzkes model, the best fit to the data for Cu films (within a TaN/Cu/TaN stack) indicated that both surface ($p = 0.05$) and grain boundary scattering ($R = 0.22$) contribute to the thin film resistivity. The values of $p$ and $R$ fall well within the range of published values \cite{7,8,Maitre,Marom,Josell,Sun1,Graham,Sun2,Chawla1,Chawla3}. Moreover, they are in good agreement with a recent review \cite{Josell} that concluded that the scattering at TaN/Cu interfaces is highly diffuse, in agreement with our results. 

By contrast, fitted grain boundary reflection coefficients for Ru and Ir were larger than for Cu with $R = 0.43$ to 0.58 for the different Ru stacks and $R = 0.47$ for Ir/SiO$_2$. Although grain boundary configurations, in particular the average misorientation angle of contiguous grains, can have an influence on the grain boundary resistance, as discussed below, a simple model of the material dependence of $R$ for polycrystalline films has been proposed by relating $R$ to the surface energy (and to the melting point) of the material \cite{Zhu}. This is fully consistent with our observations that $R$ was larger for the more refractory Ru and Ir than for Cu. Moreover, our fitted values of $R \sim 0.5$ for Ru and $R = 0.47$ for Ir are in reasonable quantitative agreement with the predictions of this model of $R = 0.55$ for Ru and $R = 0.57$ for Ir \cite{Zhu}. Note that the predicted value for Cu is $R = 0.35$, also in reasonable agreement with our results. Recently, Lanzillo calculated $R$ for twin boundaries in PGMs (Pt, Rh, Ir and Pd) by \emph{ab initio} methods \cite{Lanzillo} and found them to be higher than for Cu, in qualitative agreement with our results. It should however be noted that all such fitted $R$ values describe ''effective`` grain boundary reflection coefficients since the grain structures of the films certainly contain many different grain boundary structures. Moreover, the grain boundary transmission might also be affected by confinement effects for the thinnest films. For this reason, the quantitative understanding of grain boundary reflection coefficients both in Cu and PGMs will still require further work.

The fitted grain boundary coefficients of Ru/SiO$_2$ showed a significant reduction upon annealing ($R = 0.43$ \textit{vs.} $R = 0.58$). This may be attributed to a reduction of the average misorientation of adjacent grains, as typically observed during recrystallization processes due to the preferred movement of large-angle grain boundaries \cite{Doherty,recryst}. It has been both calculated \cite{Cesar1} and experimentally observed \cite{Lu_Science,Kim_GG_resistance,Kim_GG_resistance2} for Cu that the resistance of boundaries between randomly oriented grains is much larger than that of coherent or coincidence grain boundaries. We speculate that a similar behavior also applies to Ru grain boundaries, leading to a reduction of $R$ upon annealing. The intermediate fitted value for the TaN/Ru/TaN stack is also in qualitative agreement with this argument since the observed small grain size even after annealing may be correlated with larger grain boundary resistances than for annealed large grain Ru on SiO$_2$.

Interestingly, the best fits indicated that both Ru and Ir on SiO$_2$ showed nearly specular surface scattering with $p > 0.9$. Hence, in those films, the Mayadas--Shatzkes model suggests that, despite their small thicknesses, surface scattering did not appear to contribute strongly to the resistivity. It has been calculated that the surface scattering coefficient for a given interface should be a strong function of both the magnitude (RMS) as well as its lateral correlation length of the surface roughness \cite{Moors,FDD,Ke,RB}. Although the Ru/SiO$_2$ and Ir/SiO$_2$ films were somewhat smoother than the TaN/Cu/TaN films (Fig.~\ref{AFM}), the difference is small and the lateral correlations lengths are similar. Therefore, it appears unlikely that the difference between Ru/SiO$_2$ as well as Ir/SiO$_2$ and TaN/Cu/TaN was only due to differences in the physical surface properties. Thus, the electronic structure and the scattering potentials at the interface may contribute significantly \cite{Zahid}.

Moreover, the fits indicated strongly diffuse scattering at Ru/TaN interfaces (as in TaN/Ru/TaN stacks) with $p = 0.01$. This implies that surface scattering depends less on the conducting metal (Ru \textit{vs.} Cu) than on the cladding material of the thin film (SiO$_2$/air \textit{vs.} TaN). Note that surface roughnesses for TaN/Ru/TaN and Ru/SiO$_2$ were almost identical. Similar observations have been made for Cu \cite{7,Chawla_O2,Zheng}. In particular, Rossnagel and Kuan \cite{7} have observed that the surface scattering contribution to the thin film resistivity was lower in contact with oxides (SiO$_2$, Ta$_2$O$_5$) than with TaN, very similar to our observations. 

Several models for the surface specularity parameter $p$ have been reported in the literature \cite{FDD,Ke,RB,Soffer,FD_Ta_Cu,Zahid} that quantitatively link $p$ to the surface roughness and the intrinsic properties of the Fermi surface of the conducting material. However, only very few studies have considered the effect of the cladding material \cite{FD_Ta_Cu,Zahid}, which appears essential in view of the experimental results. Zahid \textit{et al.} \cite{Zahid} have studied the resistivity of Cu films surrounded by different metals using \textit{ab initio} calculations and found that metals can both lower as well as increase $p$ with respect to a free Cu surface, depending on the difference of the density of states at the Fermi level of conducting metal (Cu) and the cladding atom at the interface. Although the density of states at the Ru/TaN interface has not yet been calculated, the bulk densities of states of Ru and TaN at the respective Fermi levels are rather similar. Additional work is thus needed to clarify the contributions of the properties of the conductor and the cladding material and its interface on the surface scattering parameter.

\subsection{Relative contributions of surface and grain boundary scattering in the Mayadas--Shatzkes model and deviations from Matthiessen's rule}

\begin{figure*}[p]
\includegraphics[width=17cm]{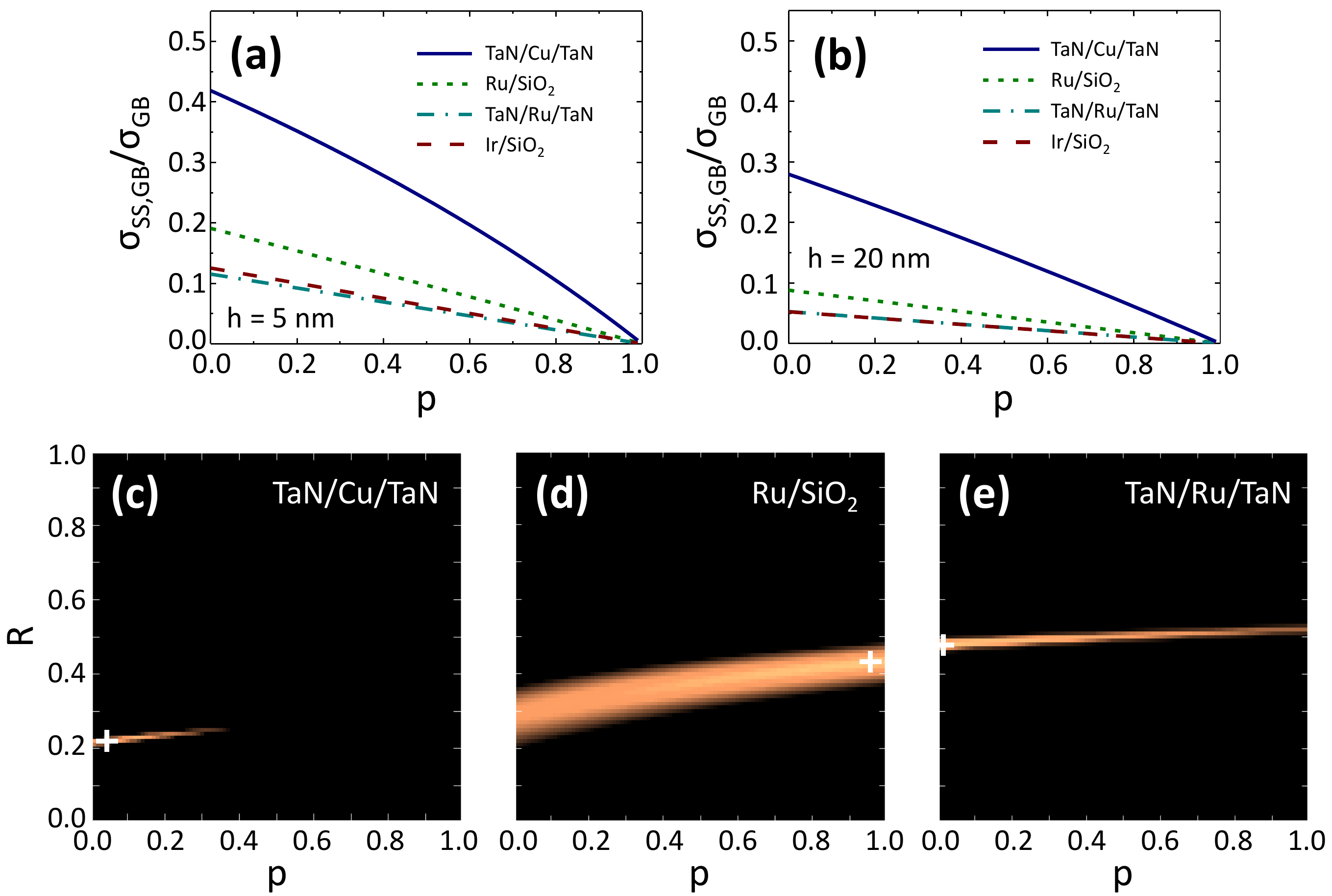} 
\caption{\label{sigmas} $\sigma_\mathrm{SS,GB}/\sigma_\mathrm{GB}$ as a function of the surface scattering parameter $p$ for (annealed) stacks as indicated for film thicknesses of 5\,nm (a) and 20\,nm (b), respectively. Experimental mean linear intercept lengths and surface scattering parameters $R$ corresponding to best fits were used. (c) -- (e) Sum of squared errors (SSE) of fits to the experimental data (\textit{cf.} Fig.~\ref{Fits}) \textit{vs.} $p$ and $R$ fitting parameters for (c) TaN/Cu/TaN, (d) Ru/SiO$_2$, and (e) TaN/Ru/TaN, all after post-deposition annealing at 420\,\deg C. The color scale corresponds to the range between $1\times$ and $4\times$ the minimum SSE for all graphs. The white crosses represent the positions of minimum SSE. }
\end{figure*}

Equation (\ref{MSS}) does not fulfill Matthiessen's rule and the contributions of grain boundary and surface scattering can therefore not be separated. While the first term in Eq.~(\ref{MSS}), $1/\rho_\mathrm{GB} \equiv \sigma_\mathrm{GB}$, describes grain boundary scattering independently of surface scattering, the second term, $1/\rho_{SS,GB} \equiv \sigma_{SS,GB}$ describes combined effects of surface and grain boundary scattering. Nonetheless, the ratio of the two terms can be evaluated and allows to shed some light on the relative importance of grain boundary and surface scattering within the Mayadas--Shatzkes model. Figure~\ref{sigmas} shows the ratio of $\sigma_{SS,GB}$ and $\sigma_\mathrm{GB}$ as a function of the surface scattering parameter $p$ for TaN/Cu/TaN, Ru/SiO$_2$, TaN/Ru/TaN, as well as Ir/SiO$_2$ and film thicknesses of 5\,nm [Fig.~\ref{sigmas}(a)] and 20\,nm [Fig.~\ref{sigmas}(b)]. Experimental mean linear intercept lengths and surface scattering parameters $R$ corresponding to best fits were used. It should be noted that in Eq.~(\ref{MSS}), a ratio of $\sigma_\mathrm{SS,GB}/\sigma_\mathrm{GB} = 0.5$ corresponds to equal contributions of the two terms. If $\sigma_\mathrm{GB} \gg \sigma_{SS,GB}$ for all values of $p$, the second term can be neglected and the thin film resistivity in the Mayadas--Shatzkes model is dominated by grain boundary scattering. However, due to the violation of Matthiessen's rule, the opposite conclusion, namely the dominance of surface scattering for $\sigma_\mathrm{GB} \approx \sigma_{SS,GB}$, is not necessarily valid.

The data indicate a general prevalence of grain boundary scattering over surface scattering within the Mayadas--Shatzkes model for all stacks even for the most diffusive case of $p = 0$ since generally $\sigma_\mathrm{SS,GB}/\sigma_\mathrm{GB} \ll 0.5$. Only for TaN/Cu/TaN (in particular for 5\,nm film thickness), $\sigma_\mathrm{SS,GB}$ contributed strongly to the overall conductivity. By contrast, the contributions were weak for PGM containing stacks---even for fully diffusive surface scattering, as observed for TaN/Ru/TaN.  

The different magnitudes of $1/\rho_\mathrm{GB} \equiv \sigma_\mathrm{GB}$ and $1/\rho_\mathrm{SS,GB} \equiv \sigma_\mathrm{SS,GB}$ have strong repercussions on the accuracy of the extracted $p$ and $R$ values. Figures~\ref{sigmas}(c)--(e) show the sum of squared errors (SSE) of the different fits as a function of the fitting parameters $p$ and $R$ for TaN/Cu/TaN, Ru/SiO$_2$, and TaN/Ru/TaN. In general, due to the small contribution of $\sigma_\mathrm{SS,GB}$, SSE minima were rather elongated along the $p$-axis but well defined along the $R$-axis. Generally, a rather weak gradient was visible along the elongated SSE minima towards the values reported in Tab.~\ref{Tab1}. Correlations between $p$ and $R$ were also visible that increase the otherwise very small errors in $R$. Nonetheless, this resulted in much larger error bars (by about $3\times$) of $p$ as compared to $R$.

This shows that the surface scattering specularity parameter of PGMs can therefore only approximately be determined by modeling of the thickness dependence of the resistivity within the Mayadas--Shatzkes model, at least for the film thicknesses and grain sizes considered here. However, the discussion above suggests that grain boundary scattering dominates the Ru and Ir thin film resistivities, even more so than for Cu, due to the much smaller MFP and that this holds independently of the exact value of $p$. This also indicates that the absolute value of $p$ should not necessarily be taken as a measure whether surface scattering contributes significantly or not.

By contrast, the contribution of $\sigma_\mathrm{SS,GB}$ was much larger for TaN/Cu/TaN [Figs.~\ref{sigmas}(a) and (b)]. This can be linked to the long MFP of 40.6\,nm and indicates that surface scattering cannot be simply neglected for thin Cu. Although both $\sigma_\mathrm{GB}$ and $\sigma_\mathrm{SS,GB}$ are reduced with increasing MFP [via the dimensionless parameters $\alpha$ and $\kappa$ in Eq.~(\ref{MSS})], $\sigma_\mathrm{SS,GB}$ appears more sensitive than $\sigma_\mathrm{GB}$, leading to an increasing prevalence of $\sigma_\mathrm{GB}$ for large $\alpha$. In addition, the grain boundary reflection coefficient of Cu, $R = 0.22$ was found to be much smaller than for PGMs ($R \sim 0.5$), which also leads to a weaker relative contribution of $\sigma_\mathrm{GB}$ for Cu with respect to PGMs.

\begin{figure}[p]
\includegraphics[width=8.5cm]{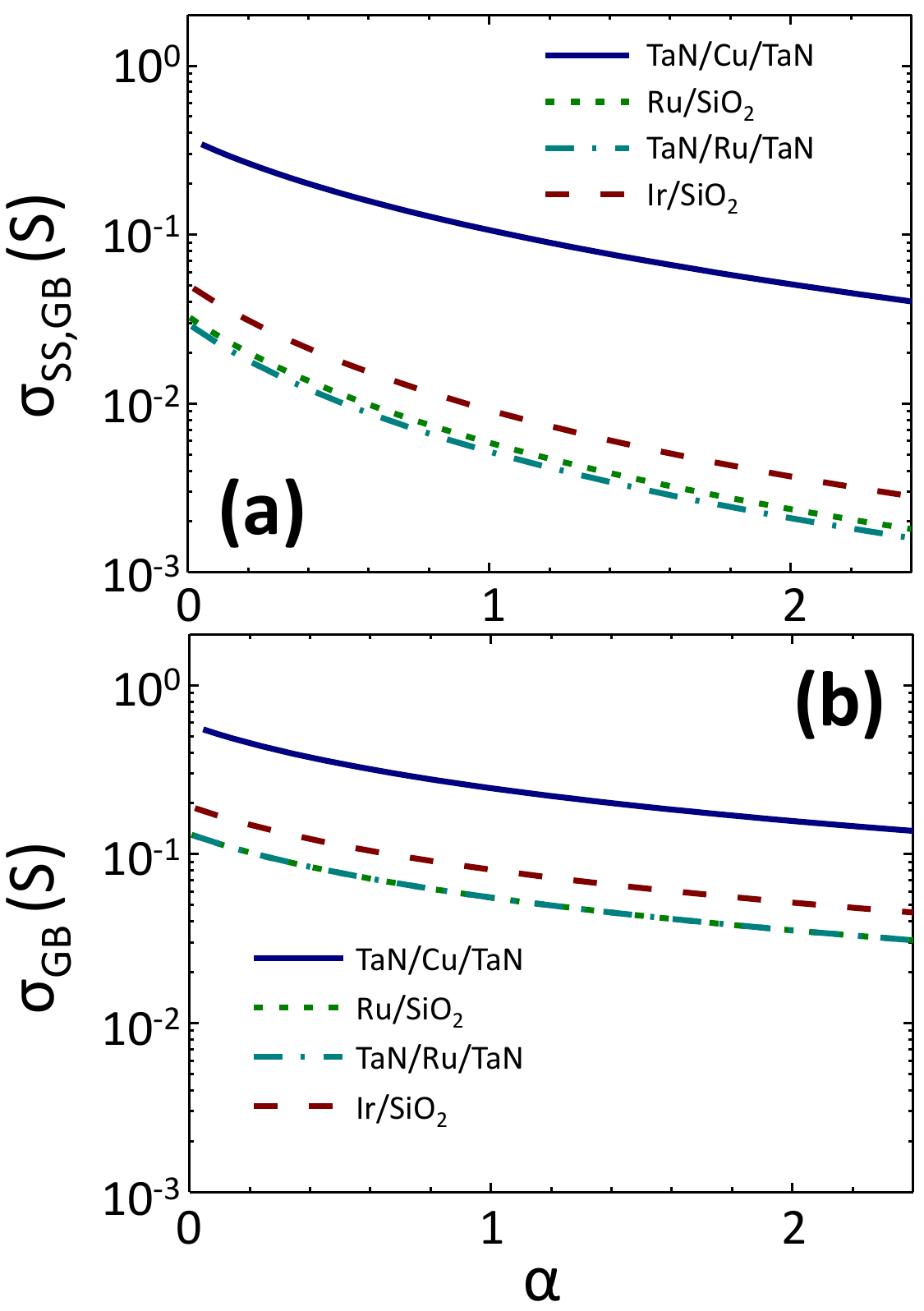} 
\caption{\label{Matthiessen} Deviations from Matthiessen's rule: (a) $\sigma_\mathrm{SS,GB}$ and (b) $\sigma_\mathrm{GB}$ as a function of the dimensionless grain boundary scattering parameter $\alpha = \left(\lambda /l\right) \times 2R\left(1-R\right)^{-1}$. Here, the thickness was set to 10 nm and fully diffuse surface scattering with $ p = 0$ was assumed to make the curves more comparable.}
\end{figure}

Among the PGMs, $\sigma_\mathrm{SS,GB}/\sigma_\mathrm{GB}$ of Ru/SiO$_2$ showed a much stronger dependence on $p$ [Figs.~\ref{sigmas}(a) and (b)] than TaN/Ru/TaN or Ir/SiO$_2$. This can be linked to deviations from Matthiessen's rule, as shown in Fig.~\ref{Matthiessen}. As pointed out by Mayadas and Shatzkes \cite{4}, the (effective) MFP that determines surface scattering in a polycrystal (\textit{i.e.} in presence of grain boundary scattering) is reduced over the bulk value by $\lambda_\mathrm{GB} = (\rho_0/\rho_\mathrm{GB})\lambda_0$. This leads to a dependence of surface scattering on $\alpha$. The effect is illustrated in Fig.~\ref{Matthiessen}(a) for $h = 10$\,nm, which shows $\sigma_\mathrm{SS,GB}$ as a function $\alpha$. To make the curves more comparable, $p = 0$ was assumed in all cases. For comparison, the dependence of $\sigma_\mathrm{GB}$ on $\alpha$ is also shown in Fig.~\ref{Matthiessen}(b). The data show that an increase of $\alpha$ (\textit{i.e.} stronger grain boundary scattering) leads to an decrease in $\sigma_\mathrm{SS,GB}$ that is generally faster than for $\sigma_\mathrm{GB}$. At $h = 10$\,nm, due to the combination of large grains and short MFP, $\alpha = 0.5$ for Ru/SiO$_2$, much smaller than for TaN/Ru/TaN ($\alpha = 1.4$), and Ir/SiO$_2$ ($\alpha = 1.6$), which leads to a relatively larger $\sigma_\mathrm{SS,GB}$ of Ru/SiO$_2$ for a given value of $p$. In practice however, the scattering at Ru/SiO$_2$ and Ru/air interfaces was found to be nearly specular and the contribution of $\sigma_\mathrm{SS,GB}$ to the thin film resistivity was also negligible for Ru/SiO$_2$. 

\subsection{Influence of the mean free path on the thickness dependence of the resistivity}

Finally, we evaluate within the Mayadas--Shatzkes model the relative impact of the different material parameters ($\lambda$, $p$, $R$, $l$) on the slope of the thickness dependence of the thin film resistivity. It has been previously proposed that metals with a shorter MFP may show a much weaker thickness dependence of their thin film resistivity. However, this effect may potentially be complemented or domineered by other factors such as the material (stack) dependence of surface and grain boundary scattering coefficients as well as the thickness dependence of the mean linear grain boundary intercept length, which will in generally depend both on the material and the applied thermal budget. 

\begin{figure}[p]
\includegraphics[width=8.5cm]{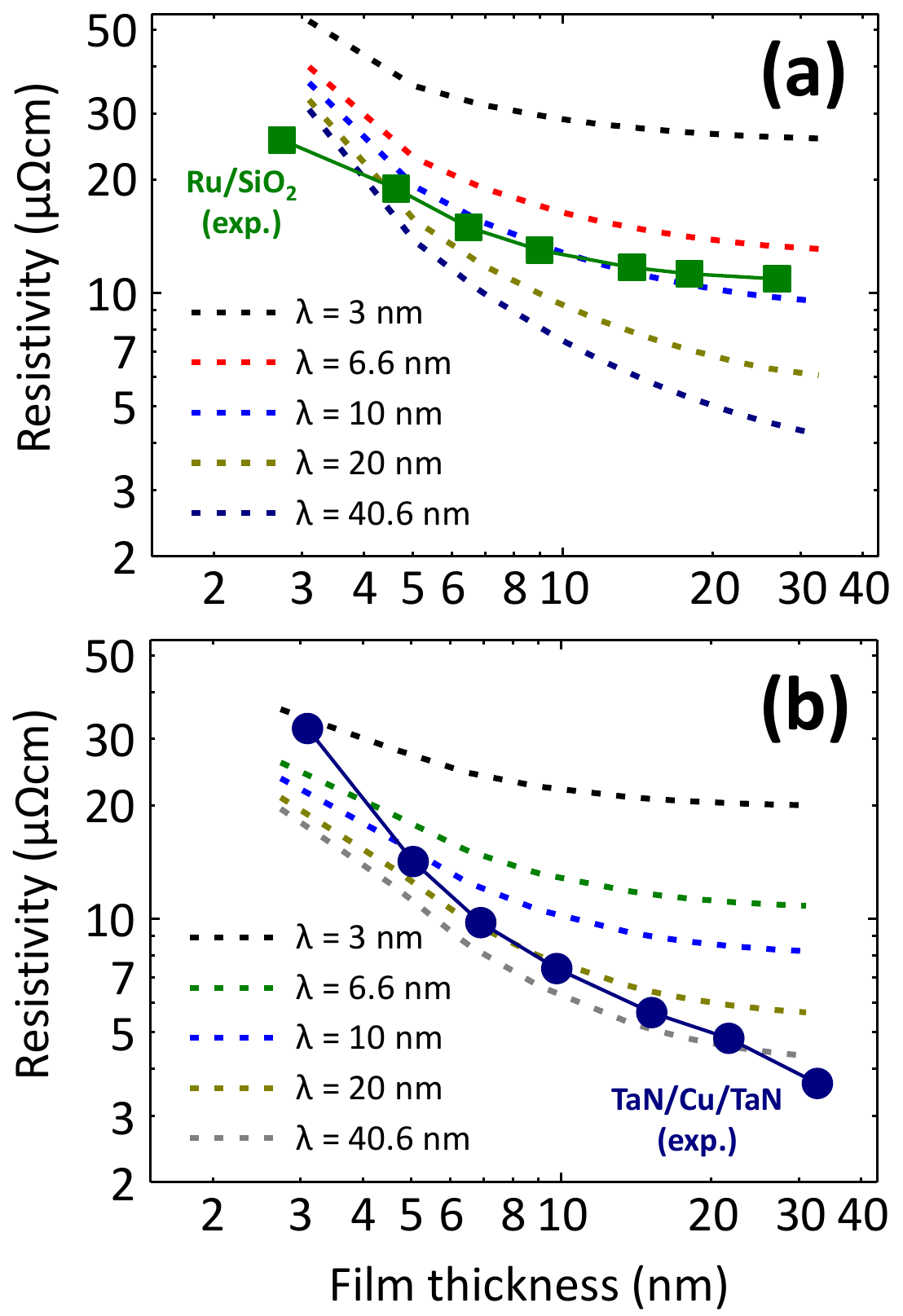} 
\caption{\label{IMFP_Calc} Calculated thickness dependence of the resistivity as a function of the mean free path (MFP) $\lambda$ with $\lambda\times\rho_0$ constant using (a) Cu parameters (\emph{i.e.} $\lambda\times\rho_0$, $p$, and $R$) and linear distances between grain boundaries as well as (b) Ru parameters and linear distances between grain boundaries. For $\lambda$ equal to the value for Ru (6.6\,nm), the simulated curve using Cu parameters in (a) is close to the Ru/SiO$_2$ experimental resistivities (green squares); analogously, the $\lambda$ of Cu (40.6\,nm) in combination with Ru parameters in (b) leads to a thickness dependence of the resistivity close to that of Cu (blue circles). This indicates that the weaker film thickness dependence of the resistivity of platinum-group metals as compared to Cu can be attributed mainly to their shorter MFPs.}
\end{figure}

To gain further insight in the importance of the electron MFP, we have calculated the expected thickness dependence of the resistivity of Cu or Ru as a function of the MFP, keeping $\lambda\times\rho_0$ constant as it is only a function of the Fermi surface morphology. The result, using the experimentally deduced parameters ($p$, $R$, and mean linear grain boundary intercept length) for Cu, is shown in Fig.~\ref{IMFP_Calc}(a). The data indicate that the overall slope of the resistivity \textit{vs.} thickness curves shows a strong dependence on the MFP. Reducing the MFP to that of Ru (6.6\,nm, see Tab.~\ref{Tab_Ab_Initio}) while keeping $\lambda\times\rho_0$ constant ($6.9\times 10^{-16}$\,$\Omega$m$^2$) leads to both a slope and absolute resistivities that are close to what was experimentally observed for annealed Ru/SiO$_2$ [Fig.~\ref{IMFP_Calc}(a)]. 

Conversely, as shown in Fig.~\ref{IMFP_Calc}(b), using the parameters obtained for annealed Ru/SiO$_2$ ($p$, $R$, and average linear intercept between grain boundaries) and increasing the MFP to that of Cu (40.6\,nm, see Tab.~\ref{Tab_Ab_Initio}) while again keeping $\lambda\times\rho_0$ constant ($5.0\times 10^{-16}$\,$\Omega$m$^2$) leads to a slope almost identical to that experimentally observed for TaN/Cu/TaN. The residual differences stem from the material dependence of $\lambda\times\rho_0$ (see Tab.~\ref{Tab_Ab_Initio}), $R$, and $p$ (see Tab.~\ref{Tab1}), as well as from the different thickness dependence of the mean linear grain boundary intercept length, and are rather small. The larger deviations for the 3\,nm thick films in both graphs can be ascribed within the Mayadas--Shatzkes model to the much stronger contribution of surface scattering to the Cu resistivity, which becomes significant only for such small thicknesses. As a whole, however, this confirms that the shorter MFP is the main root cause for the different thickness dependence of the resistivity of Cu and the PGMs. 

\section{Conclusion}

In conclusion, we have studied the thickness dependence of the resistivity of ultrathin PGM films in the range between 3 and 30\,nm. All studied PGMs (Ru, Pd, Ir, Pt) show a much weaker thickness dependence than Cu, the reference material. As a consequence, PGM thin film show comparable or even lower resistivities than Cu for film thicknesses of about 5\,nm and below. 

The thickness dependence of the resistivity of TaN/Cu/TaN, Ru/SiO$_2$, TaN/Ru/TaN, and Ir/SiO$_2$ was modeled using the Mayadas--Shatzkes model \cite{4} and  experimentally determined mean linear grain boundary intercept lengths as well as \emph{ab initio} calculations of the MFP for bulk metals. Fitted grain boundary scattering coefficients for Ru and Ir ($R \sim 0.4$ to 0.5) were significantly higher than for Cu ($R = 0.22$), in good qualitative agreement with recent calculations \cite{Zhu,Lanzillo}. The model also found nearly specular scattering ($p > 0.9$) was observed for both Ru and Ir on SiO$_2$ but the interface scattering was much more diffuse ($p \approx 0$) for TaN/Ru/TaN indicating that specular surface scattering is not an intrinsic material property of Ru. This behavior is currently not yet well understood owing to the lack of a general predictive theory for the material dependence of interface scattering. However, it should be noted that in all cases---irrespective of $p$---surface scattering contributed only weakly to the overall resistivity, which was dominated by grain boundary scattering, except for the thinnest TaN/Cu/TaN films.

Simulations within the Mayadas--Shatzkes model showed that the much shorter MFP of Ru and Ir was indeed responsible for the much weaker thickness dependence of the thin film resistivity. This confirms earlier predictions \cite{5,Zhang,Naeemi} and justifies the usage of $(\lambda\rho_0)^{-1}$ as a figure of merit of alternative metals for beyond-Cu interconnects \cite{13,14}, in particular with respect to the the expected scaling behavior. Indeed, PGMs---and in particular Ru---have recently shown excellent prospects to replace Cu in future nanoscale interconnects with scaled widths of 10\,nm and below \cite{ACSRuIC,AM1,AM2,AM3,DuttaEDL}.

\section*{Supplementary Material}

See the supplementary material for the derivation of the correct definition of $\alpha$ in Eq.~(\ref{MSS}).

\begin{acknowledgments}

The authors would like to thank Sofie Mertens, Thomas Witters, and Karl Opsomer (imec) for the support of the PVD depositions, as well as Christian Witt (GlobalFoundries) for many stimulating discussions. Olivier Richard and Niels Bosman are acknowledged for the TEM imaging as well as Danielle Vanhaeren, Lien Landeloos, and Inge Vaesen (imec) for the AFM measurements. Johan Meersschaut is acknowledged for the RBS measurements. S.D. would like to thank Anamul Hoque, Kristof Peeters, Michiel Vandemaele, Christopher Gray, and Margo Billen (KU Leuven) for their assistance in the TEM image analysis. This work has been supported by imec's industrial affiliate program on nano-interconnects.

\end{acknowledgments}

\clearpage

\begin{table*}[p]
\caption{\label{Tab_Ab_Initio} Product of the relaxation time $\tau$ and the resistivity $\rho$ for platinum-group metals and Cu, as determined by \emph{ab initio} calculations in combination with Eq.~(\ref{lambda}), as well as calculated relaxation times due to electron--phonon scattering, $\tau_c$. Using experimental bulk resistivities $\rho_0$ \cite{Resistivity}, relaxation times $\tau_0$ can be calculated from $\tau\times\rho$. Mean free paths $\lambda_c$ and $\lambda_0$, can then be deduced from $\tau_c$ and $\tau_0$, respectively, using the Fermi velocity $v_\mathrm{ave}$ averaged over transport directions perpendicular to [001] for hcp metals (Ru, Os) or [111] for fcc metals (Cu, Rh, Pd, Ir, Pt). $(\lambda_0\rho_0)^{-1}$ can be considered as a figure of merit for the resistivity scaling to small dimensions.}
\begin{tabular}{lcccccccc}
\hline\hline 
& $\tau \times\rho$  & $\tau_c$ & $\rho_0$ & $\tau_0$ & $v_\mathrm{ave}$ & $\lambda_\mathrm{c}$ & $\lambda_0$ & $(\lambda_0\rho_0)^{-1}$ \\ 
& ($10^{-22}$\,$\Omega$ms) & (fs) & ($\mu\Omega$\,cm) & (fs) & ($10^5$\,m/s) & (nm) & (nm) & ($10^{15}$\,S/m$^2$)\\
\hline 
Cu & 6.36 & 23.2 & 1.71 & 37.2 & 10.92 & 25.3 & 40.6 & 1.4\\ 

Ru & 6.68 & 8.6 & 7.6 & 8.8 & 7.47 & 6.4 & 6.6 & 2.0 \\ 

Rh & 5.18 & 10.3 & 4.8 & 10.8 & 6.92 & 7.1 & 7.5 & 2.8 \\ 

Pd & 11.00 & 19.9 & 10.7 & 10.3 & 3.18 & 6.3 & 3.3 & 2.8\\ 

Ir & 4.77& 2.7 & 5.2 & 9.2 & 8.83 & 2.4 & 8.1 & 2.4 \\ 

Os & 6.81 & 9.1 & 10.0 & 6.8 & 8.39 & 7.6 & 5.7 & 1.8 \\ 

Pt & 5.64 & 8.6 & 10.6 & 5.3 & 5.20 & 4.5 & 2.8 & 3.4 \\ 

\hline\hline
\end{tabular} 
\end{table*}

\begin{table*}[p]
\caption{\label{Tab1} Best fitting parameters, $p$ and $R$, along with mean free path $\lambda_0$ and bulk resistivity $\rho_0$ \cite{Resistivity} used as input parameters in the Mayadas--Shatzkes model (\emph{cf.} Tab.~\ref{Tab_Ab_Initio}). The coefficient of determination adjusted for the number of fitting parameters, $R_\mathrm{adj}^2$, is also given for each data set.}

\begin{tabular}{lccccc}
\hline\hline 
& $p$ & $R$ & $\lambda_0$ (nm) & $\rho_0$ ($\mu\Omega$\,cm) & $R_\mathrm{adj}^2$ \\ 
\hline 
Ru/SiO$_2$ (as deposited) & 0$.93 \pm 0.08$  & $0.58 \pm 0.02$ & 6.6 & 7.6 & 0.99 \\ 

Ru/SiO$_2$ (annealed) & $0.98 \pm 0.09$  & $0.43 \pm 0.04$ & 6.6 & 7.6 & 0.92 \\ 

TaN/Ru/TaN (annealed)& $0.01 \pm 0.06$ & $0.48 \pm 0.02$ & 6.6 & 7.6 & 0.99 \\ 

Ir/SiO$_2$ (annealed) & $0.94 \pm 0.09$ & $0.47 \pm 0.03$ & 8.1 & 5.2 & 0.95 \\ 

TaN/Cu/TaN (annealed) & $0.05 \pm 0.04$ & $0.22 \pm 0.02$ & 40.6 & 1.71 & 0.99 \\ 

\hline\hline
\end{tabular} 
\end{table*}

\clearpage

\begin{thebibliography}{(99}
\bibitem{1}	K. Fuchs, Math. Proc. Camb. Philos. Soc. \textbf{34}, 100 (1938).

\bibitem{2}	E. H. Sondheimer, Adv. Phys. \textbf{1}, 1 (1952).

\bibitem{ZB} X.-G. Zhang and W. H. Butler, Phys. Rev. B \textbf{51}, 10085 (1995).

\bibitem{3}	A. F. Mayadas, M. Shatzkes, and J. F. Janak, Appl. Phys. Lett. \textbf{14}, 345 (1969).

\bibitem{4}	A. F. Mayadas and M. Shatzkes, Phys. Rev. B \textbf{1}, 1382 (1970).

\bibitem{Tesa} Z. Te\v{s}anovi\'{c}, M. V. Jari\'{c}, and S. Maekawa, Phys. Rev. Lett. \textbf{57}, 2760 (1986).

\bibitem{NA} N. Trivedi and N. W. Ashcroft, Phys. Rev. B \textbf{38}, 12298 (1988).

\bibitem{Mey} A. E. Meyerovich and I. V. Ponomarev, Phys. Rev. \textbf{65}, 155413 (2002).

\bibitem{Moors} K. Moors, B. Sor\'{e}e, and W. Magnus, J. Appl. Phys. \textbf{118}, 124307 (2015).

\bibitem{Hedge1} G. Hegde, M. Povolotskyi, T. Kubis, J. Charles, and G. Klimeck, J. Appl. Phys. \textbf{115}, 123704 (2014).

\bibitem{Jones} S. L. T. Jones, A. Sanchez-Soares, J. J. Plombon, A. P. Kaushik, R. E. Nagle, J. S. Clarke, and J. C. Greer, Phys. Rev. B \textbf{92}, (2015).

\bibitem{Munoz} R. C. Munoz and C. Arenas, Appl. Phys. Rev. \textbf{4}, 011102 (2017).

\bibitem{Lanzillo} N. A. Lanzillo, J. Appl. Phys. \textbf{121}, 175104 (2017).

\bibitem{7}	S. M. Rossnagel and T. S. Kuan, J. Vac. Sci. Technol. B \textbf{22}, 240 (2004).

\bibitem{8}	J. J. Plombon, E. Andideh, V. M. Dubin, and J. Maiz, Appl. Phys. Lett. \textbf{89}, 113124 (2006).

\bibitem{Maitre} S. Ma\^{i}trejean, R. Gers, T. Mourier, A. Toffoli, and G. Passemard, Microelectron. Engin. \textbf{83}, 2396 (2006).

\bibitem{Marom} H. Marom, J. Mullin, and M. Eizenberg, Phys. Rev. B \textbf{74}, 045411 (2006).

\bibitem{Josell} D. Josell, S. H. Brongersma, and Z. T\H{o}kei, Annu. Rev. Mater. Res. \textbf{39}, 231 (2009).

\bibitem{Sun1} T. Sun, B. Yao, A. P. Warren, K. Barmak, M. F. Toney, R. E. Peale, and K. R. Coffey, Phys. Rev. B \textbf{79}, 041402(R) (2009).

\bibitem{Graham} R. L. Graham, G. B. Alers, T. Mountsier, N. Shamma, S. Dhuey, S. Cabrini, R. H. Geiss, D. T. Read, and S. Peddeti, Appl. Phys. Lett. \textbf{96}, 042116 (2010).

\bibitem{Sun2} T. Sun, B. Yao, A. P. Warren, K. Barmak, M. F. Toney, R. E. Peale, and K. R. Coffey, Phys. Rev. B \textbf{81}, 155454 (2010).

\bibitem{Chawla1} J. S. Chawla, F. Gstrein, K. P. O'Brien, J. S. Clarke, and D. Gall, Phys. Rev. B \textbf{84}, 235423 (2011).

\bibitem{Chawla3} J. S. Chawla and D. Gall, Appl. Phys. Lett. \textbf{94}, 252101 (2009).

\bibitem{DeVries} J. W. C. De Vries, Thin Solid Films \textbf{167}, 25 (1988).

\bibitem{Tay} M. Tay, K. Li, and Y. Wu, J. Vac. Sci. Technol. B \textbf{23}, 1412 (2005).

\bibitem{Camacho} J. M. Camacho and A. I. Oliva, Microelectron. J. \textbf{36}, 555 (2005).

\bibitem{5}	P. Kapur, J. P. McVittie, and K. C. Saraswat, IEEE Trans. Electron Devices \textbf{49}, 590 (2002).

\bibitem{6} W. Steinh\"{o}gl, G. Schindler, G. Steinlesberger, and M. Engelhardt, Phys. Rev. B \textbf{66}, 075414 (2002).

\bibitem{9} F. Chen and D. Gardner, IEEE Electron Device Lett. \textbf{19}, 508 (1998).

\bibitem{10} P. Kapur, G. Chandra, J. P. McVittie, and K. C. Saraswat, IEEE Trans. Electron Devices \textbf{49}, 598 (2002).

\bibitem{11} K. J. Kuhn, IEEE Trans. Electron Devices \textbf{59}, 1813 (2012).

\bibitem{Ceyhan} A. Ceyhan and A. Naeemi, IEEE Trans. Electron Devices \textbf{60}, 4041 (2013).

\bibitem{12} C. Adelmann, L. G. Wen, A. P. Peter, Y. K. Siew, K. Croes, J. Swerts, M. Popovici, K. Sankaran, G. Pourtois, S. Van Elshocht, J. B\"{o}mmels, and Z. T\H{o}kei, Proc. IEEE Int. Interconnect Technol. Conf., pp. 173 (2014).

\bibitem{13} K. Sankaran, S. Clima, M. Mees, C. Adelmann, Z. T\H{o}kei, and G. Pourtois, Proc. IEEE Int. Interconnect Technol. Conf. pp. 193 (2014).

\bibitem{Kirou2} K. Sankaran, S. Clima, M. Mees, and G. Pourtois, ECS J. Solid State Sci. Technol. \textbf{4}, N3127 (2015).

\bibitem{14} D. Gall, J. Appl. Phys. \textbf{119}, 85101 (2016).

\bibitem{Naeemi} C. Pan and A. Naeemi, IEEE Electron Device Lett. \textbf{35}, 250 (2013).

\bibitem{Zhang} W. Zhang, S. H. Brongersma, O. Richard, B. Brijs, R. Palmans, L. Froyen, and K. Maex, Microelectron. Engin. \textbf{76}, 146 (2004).

\bibitem{Simbeck} A. J. Simbeck, N. Lanzillo, N. Kharche, M. J. Verstraete, and S. K. Nayak, ACS Nano \textbf{6}, 10449 (2012).

\bibitem{Gao} N. Gao, J. C. Li, and Q. Jiang, Appl. Phys. Lett. \textbf{103}, 263108 (2013).

\bibitem{ACSRuIC} L. G. Wen, P. Roussel, O. Varela Pedreira, B. Briggs, B. Groven, S. Dutta, M. I. Popovici, N. Heylen, I. Ciofi, K. Vanstreels†, F. W. \O sterberg, O. Hansen, D. H. Petersen, K. Opsomer, C. Detavernier, C. J. Wilson, S. Van Elshocht, K. Croes, J. B\"ommels, Z. T\H{o}kei, and C. Adelmann, ACS Appl. Mater. Interfaces \textbf{8}, 26119 (2016).

\bibitem{ASTM} ASTM E112-13, Standard Test Methods for Determining Average Grain Size, available at \texttt{www.astm.org} (ASTM International, West Conshohocken, PA, 2013).

\bibitem{Abrams} H. Abrams, Metallography \textbf{4}, 59 (1971).

\bibitem{K8} P. Giannozzi, S. Baroni, N. Bonini, M. Calandra, R. Car, C. Cavazzoni, D. Ceresoli, G. L. Chiarotti, M. Cococcioni, I. Dabo, A. Dal Corso, S. Fabris, G. Fratesi, S. de Gironcoli, R. Gebauer, U. Gerstmann, C. Gougoussis, A. Kokalj, M. Lazzeri, L. Martin-Samos, N. Marzari, F. Mauri, R. Mazzarello, S. Paolini, A. Pasquarello, L. Paulatto, C. Sbraccia, S. Scandolo, G. Sclauzero, A. P. Seitsonen, A. Smogunov, P. Umari, and R. M. Wentzcovitch, J. Phys.: Condens. Matter \textbf{21}, 395502 (2009).

\bibitem{K9} P. E. Bl\"ochl, Phys. Rev. B \textbf{50}, 17953 (1994).

\bibitem{K10} J. P. Perdew, K. Burke, and M. Ernzerhof, Phys. Rev. Lett. \textbf{77}, 3865 (1996).

\bibitem{Allen} P. B. Allen, T. P. Beaulac, F. S. Khan, W. H. Butler, F. J. Pinski, and J. C. Swihart, Phys. Rev B \textbf{34}, 4331 (1986).

\bibitem{ITRS} International Technology Roadmap for Semiconductors -- ITRS 2.0, available at \texttt{www.itrs2.net}, last accessed 10/24/2016.

\bibitem{PR} P. Roussel, I. Ciofi, R. Degraeve, V. V. Gonzalez, N. Jourdan, R. Baert, D. Linten, J. B\"ommels, G. Groeseneken, and A. Thean, Proc. IEEE Intern. Reliability Phys. Symp., IT-2 (2016).

\bibitem{Jacoboni} C. Jacoboni, \textit{Theory of Electron Transport in Semiconductors} (Springer, Berlin, Heidelberg, 2010).

\bibitem{Mahan} G. D. Mahan, \textit{Many-Particle Physics} (Springer, Boston, 2000).

\bibitem{Resistivity} J. Bass, Electrical Resistivity, Kondo and Spin Fluctuation Systems, Spin Glasses and Thermopower, in \textit{Landolt-B{\"o}rnstein -- Group III Condensed Matter}, edited by K.-H. Hellwege and L. J. Olsen, Vol. 15A, SpringerMaterials, available at {\tt dx.doi.org/10.1007/b29240} (Springer Verlag, Berlin, Heidelberg, 1983). 

\bibitem{Ru_aniso} E. M. Savitskii, P. V. Gel'd, V. E. Zinov'ev, N. B. Gorina, and V. P. Polyakova, Sov. Phys. Dokl. \textbf{21}, 456 (1976); N. V. Volkenshteyn, V. Ye. Startsev, V. I. Cherepanov, V. M. Azhazha, G. P. Kovtun, and V. A. Yelenskiy, Phys. Met. Metallogr \textbf{45}, 54 (1978).

\bibitem{Zhou} Y. Zhou, S. Sreekala, P. M. Ajayan, and S. K. Nayak, J. Phys.: Condens. Matter \textbf{20}, 095209 (2008).

\bibitem{Hedge2} G. Hegde, R. C. Bowen, and M. S. Rodder, Appl. Phys. Lett. \textbf{109}, 193106 (2016).

\bibitem{RemGS} Although Mayadas and Shatzkes in Ref. \cite{4} employ the term \textit{grain size}, their derivation of Eq.~(\ref{MSS}) is actually based on the average linear distance between grain boundaries, \textit{i.e.} the mean linear grain boundary intercept length. While average grain size and mean linear intercept length are typically found to be proportional to each other, the proportionality constant deviates significantly from unity and depends on the grain size distribution. See \textit{e.g.}, A. Thorvaldsen, Acta Materialia \textbf{45}, 595 (1997). 

\bibitem{RemR} The definition of $\alpha$ in Mayadas and Shatzkes, Ref. \cite{4}, misses a factor of 2. For the derivation of the correct definition of $\alpha$, see the supplementary material. Our results can be compared to studies based on the original definition of $\alpha$ by using $R_\textrm{MS} = 2R/(1+R)$.

\bibitem{Soffer} S. B. Soffer, J. Appl. Phys. \textbf{38}, 1710 (1967).

\bibitem{Marom_GS} H. Marom, M. Ritterband, and M. Eizenberg, Thin Solid Films \textbf{510}, 62 (2006).

\bibitem{Choi} D. Choi, X. Liu, P. K. Schelling, K. R. Coffey, and K. Barmak, J. Appl. Phys. \textbf{115}, 104308 (2014).

\bibitem{Zhu} Y. F. Zhu, X. Y. Lang, W. T. Zheng, and Q. Jiang, ACS Nano \textbf{4}, 3781 (2010).

\bibitem{Doherty} R. D. Doherty, D. A. Hughes, F. J. Humphreys, J. J. Jonas, D. Juul Jensen, M. E. Kassner, W. E. King, T. R. McNelley, H. J. McQueen, and A. D. Rollett, Mater. Sci. Engin. A \textbf{238}, 219 (1997).

\bibitem{recryst} F. J. Humphreys and M. Hatherly, \textit{Recrystallization and Related Annealing Phenomena}, 2$^\mathrm{nd}$ ed. (Elsevier, Oxford, 2004).

\bibitem{Cesar1} M. C\'{e}sar, D. Liu, D. Gall, and H. Guo, Phys. Rev. Applied \textbf{2}, 44007 (2014).

\bibitem{Lu_Science} L. Lu, Y. Shen, X. Chen, L. Qian, and K. Lu, Science \textbf{304}, 422 (2004).

\bibitem{Kim_GG_resistance} T.-H. Kim, X.-G. Zhang, D. M. Nicholson, B. M. Evans, N. S. Kulkarni, B. Radhakrishnan, E. A. Kenik, and A.-P. Li, Nano Lett. \textbf{10}, 3096 (2010).

\bibitem{Kim_GG_resistance2} T.-H. Kim, D. M. Nicholson, X.-G. Zhang, B. M. Evans, N. S. Kulkarni, E. A. Kenik, H. M. Meyer, B. Radhakrishnan, and A.-P. Li, Jpn. J. Appl. Phys. \textbf{50}, 08LB09 (2011).

\bibitem{FDD} B. Feldman, R. Deng, and S. T. Dunham, J. Appl. Phys. \textbf{103}, 113715 (2008).

\bibitem{Ke} Y. Ke, F. Zahid, V. Timoshevskii, K. Xia, D. Gall, and H. Guo, Phys. Rev. B \textbf{79}, 155406 (2009).

\bibitem{RB} J. M. Rickman and K. Barmak, J. Appl. Phys. \textbf{112}, 13704 (2012).

\bibitem{Zahid} F. Zahid, Y. Ke, D. Gall, and H. Guo, Phys. Rev. B \textbf{81}, 045406 (2010).

\bibitem{Chawla_O2} J. S. Chawla, F. Zahid, H. Guo, and D. Gall, Appl. Phys. Lett. \textbf{97}, 132106 (2010).

\bibitem{Zheng} P. Y. Zheng, R. P. Deng, and D. Gall, Appl. Phys. Lett. \textbf{105}, 131603 (2014).

\bibitem{FD_Ta_Cu} B. Feldman and S. T. Dunham, Appl. Phys. Lett. \textbf{95}, 222101 (2009).

\bibitem{AM1} X. Zhang, H. Huang, R. Patlolla, W. Wang, F. W. Mont, J. Li, C.-K. Hu, E. G. Liniger, P. S. McLaughlin, C. Labelle, E. T. Ryan, D. Canaperi, T. Spooner, G. Bonilla, and D. Edelstein, Proc. IEEE Int. Interconnect Technol. Conf., pp. 31 (2016).

\bibitem{AM2} L. G. Wen, C. Adelmann, O. Varela Pedreira, S. Dutta, M. Popovici, B. Briggs, N. Heylen, K. Vanstreels, C. J. Wilson, S. Van Elshocht, K. Croes, J. B\"ommels, and Z. T\H{o}kei, Proc. IEEE Int. Interconnect Technol. Conf., pp. 34 (2016).

\bibitem{AM3} J. S. Chawla, S. H. Sung, S. A. Bojarski, C. T. Carver, M. Chandhok, R. V. Chebiam, J. S. Clarke, M. Harmes, C. J. Jezewski, M. J. Kobrinski, B. J. Krist, M. Mayeh, R. Turkot, and H. J. Yoo, Proc. IEEE Int. Interconnect Technol. Conf., pp. 63 (2016).

\bibitem{DuttaEDL} S. Dutta, S. Kundu, A. Gupta, G. Jamieson, J. F. Gomez Granados, J. B\"ommels, C. J. Wilson, Z. T\H{o}kei, and C. Adelmann, IEEE Electron Device Lett. \textbf{99}, in print, DOI: 10.1109/LED.2017.2709248 (2017).

\end{thebibliography}
\end{document}